\documentclass[12pt,a4paper]{article}
\usepackage[dvips]{epsfig}
\title{
  {\vspace{-3cm} \normalsize \hfill
    \parbox{38mm}{MS-TPI-99-9 \\
                  cond-mat/9908246}  }\\[25mm]
  Analytical Calculation of the Nucleation Rate\\
  for First Order Phase Transitions\\
  beyond the Thin Wall Approximation
  }
\author{Gernot M\"unster and Sabine Rotsch%
        \thanks{present address: SAP AG, D-69190 Walldorf}\\
        Institut f\"ur Theoretische Physik I,
        Universit\"at M\"unster\\
        Wilhelm-Klemm-Str.~9, D-48149 M\"unster, Germany\\
        e-mail: munsteg@uni-muenster.de}
\date{August 16, 1999}
\newcommand{\be}{\begin{equation}}
\newcommand{\ee}{\end{equation}}
\newcommand{\bea}{\begin{eqnarray}}
\newcommand{\eea}{\end{eqnarray}}
\newcommand{\e}{\mathrm{e}}
\newcommand{\Hc}{\mathcal{H}}
\newcommand{\Mc}{\mathcal{M}}

\newcommand{\rt}{\tilde{r}}
\newcommand{\Rt}{\tilde{R}}
\newcommand{\nt}{\tilde{\eta}}
\newcommand{\Mcn}{\mathcal{M}^{(0)}}
\newcommand{\Mn}{M^{(0)}}
\newcommand{\Vn}{V^{(0)}}
\newcommand{\omn}{\omega^{(0)}}
\begin{document}
\maketitle

\begin{abstract}
First order phase transitions in general proceed via nucleation of
bubbles. A theoretical basis for the calculation of the nucleation rate
is given by the homogeneous nucleation theory of Langer and its field
theoretical version of Callan and Coleman. We have calculated the
nucleation rate beyond the thin wall approximation by expanding the
bubble solution and the fluctuation determinant in powers of the
asymmetry parameter. The result is expressed in terms of physical model
parameters.
\\[5mm]
PACS numbers: 05.70.Fh, 11.10.Kk, 64.60.Qb\\
Keywords: Phase transitions, Field theory, Nucleation theory
\end{abstract}
%
\section{Introduction}

First order phase transitions are a common phenomenon in statistical
mechanics and in field theory \cite{Stanley}. They are characterized by
the discontinuous change of an order parameter or other physical
quantities as some driving parameter, e.g.\ temperature, is varied. In
general they are associated with a latent heat. In the theory of
elementary particles different phase transitions, which play a role in
the evolution of the early universe, are predicted to be of first order.
Among them is the electroweak phase transition, which has been
investigated intensively in recent years, and the grand unification
phase transition, which might be related to the inflationary epoch of
the universe \cite{Guth}.

There are different mechanisms by which a first order phase transition
can take place. In many cases it proceeds via nucleation of
bubbles. Consider evaporation for example. If the temperature crosses
the transition point the system enters a metastable state. In this
state bubbles of the new, stable phase form spontaneously, which may
then expand and lead to the completion of the transition to the new
phase. Such metastable states have first been mentioned by Fahrenheit
\cite{Fahrenheit}. A theory of the formation of bubbles in liquid
systems has been developed by Becker and D\"oring \cite{Becker}. In
the framework of the Ginzburg-Landau theory of phase transitions a
phenomenological treatment was given by Cahn and Hilliard
\cite{Cahn}. The theory of bubble nucleation was put on a profound
theoretical basis by Langer \cite{Langer67,Langer68,Langer69}. His
approach allows for a systematical treatment of the nucleation rate. A
review is given in \cite{Gunton}. In the context of quantum field
theory the nucleation theory was developed by Voloshin et al.\
\cite{Voloshin}, Callan and Coleman \cite{Coleman,Callan} and Affleck
\cite{Affleck}. Callan and Coleman presented an approach to the decay
of an unstable vacuum in the framework of Euclidean quantum field
theory. Although developed independently, it is very similar to
Langer's formulation in terms of functional integrals. A nice
exposition is given in Coleman's book \cite{Aspects}.

The starting point of nucleation theory is the classical Ginzburg-Landau
potential for the order parameter. It has an absolute minimum,
corresponding to the stable phase, and one (or more) other minima. When
the first order phase transition is approached, a previously higher
minimum gets lower and lower, and when the transition point is crossed,
it becomes the new absolute minimum. There is a barrier between the
minima such that the system does not immediately go over into the new
minimum but remains in a metastable state. This state is stable against
small fluctuations. Due to fluctuations small regions (bubbles) of the
stable phase may form spontaneously. Their creation leads to a gain in
energy proportional to the volume,
\be
- \Hc_V = \frac{4 \pi}{3} R^3 \eta \,,
\ee
where $R$ is the radius of the bubble and $\eta$ is the difference of
the potentials between the two minima. On the other hand a surface
energy
\be
\Hc_S = 4 \pi R^2 \sigma
\ee
has to be supplied, and the total energy associated with the bubble is
approximately given by
\be
\Hc_b (R) = \Hc_S + \Hc_V
= 4 \pi R^2 (\sigma - \frac{R}{3} \eta) \,.
\ee
For small $R$ this function increases with $R$ so that small bubbles
tend to shrink back to zero. Only if the radius exceeds the
critical size of
\be
R_c = \frac{2 \sigma}{\eta} \,,
\ee
where
\be
\frac{d \Hc_b (R_c)}{d R} = 0 \,,
\ee
the bubble will expand and lead to the transformation of the metastable
phase into the stable one in the whole volume.

The process described above is called homogeneous nucleation in contrast
to heterogeneous nucleation, where impurities, like dust or ice
crystals, trigger the phase transition. The cosmological phase
transitions mentioned earlier are homogeneous. The formation of bubbles
in homogeneous nucleation theory is analogous to quantum mechanical
tunneling through a potential barrier. In fact, the description of
tunneling by means of Euclidean functional integrals leads to a
completely equivalent formalism. This fact is the basis of the relation
between Langer's work and that of Callan and Coleman.

The rate in which the phase transition proceeds is essentially
determined by the average time until a critical bubble forms
spontaneously.
A critical bubble of radius $R_c$ is a solution of the field equations
coming from the Ginzburg-Landau Hamiltonian.
It has an energy $\Hc_c = \Hc_b (R_c)$.
The nucleation rate $\Gamma$ per time and volume is proportional to the
Boltzmann factor of a critical bubble and can be written as
\be
\Gamma = \mathcal{A} \e^{- \Hc_c} \,,
\ee
which has already been found by Arrhenius \cite{Arrhenius}.

For practical applications it is important to know the prefactor $\cal
A$. Langer's theory gives an expression for $\cal A$ in terms of the
determinant of the operator of fluctuations around the critical bubble.
It is the main object of this article to calculate the nucleation rate
including the prefactor in the framework of scalar field theory, i.e.,
Ginzburg-Landau theory with fluctuations. Some elements of the
calculation have been supplied by Langer \cite{Langer67}, but a complete
analytical calculation has been missing in the literature so far. A
numerical method for the evaluation of the nucleation rate has been
presented by Baacke and Kiselev \cite{Baacke}.

In general it is not possible to find an analytical solution of the
field equations for finite potential differences $\eta$. An
approximation, where the field equations can be solved exactly, is the
``thin wall approximation'' \cite{Linde}. This is the limiting case
where $\eta$ is much smaller than the height of the potential barrier.
In this case the thickness of the wall of a critical bubble is much
smaller than its radius, and its density profile can be approximated by
a step function.

In this article we calculate the nucleation rate analytically beyond the
thin wall approximation. We do this by expanding all quantities in
powers of $\eta$ and calculating the logarithm of the functional
determinant of the fluctuation operator in terms of powers of $\eta$.
The leading term corresponds to the thin wall approximation. For the
calculation of the determinant we employ the Seeley expansion of the
associated heat kernel on the one hand, and the spectrum of the
fluctuation operator on the other hand. Ultraviolet divergencies require
renormalization as usual. The resulting expression for the nucleation
rate $\Gamma$ is expressed in terms of renormalized parameters of the
effective potential.
%
%
\section{Nucleation theory}

\begin{figure}[hbt]
\vspace{.8cm}
\centering
{\unitlength1cm
\begin{picture}(9,5)
\put(9.1,1.35){$\phi$}
\put(4.4,5.2){$U$}
\put(6.65,0.9){$\phi_+$}
\put(1.9,1.8){$\phi_-$}
\put(4.7,0.2){$\eta$}
\epsfig{file=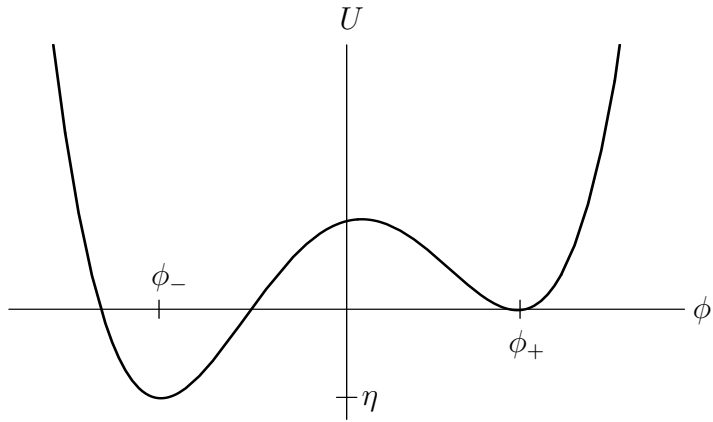,width=9cm,height=5cm}
\end{picture}
}
\parbox[t]{0.8\textwidth}{
\caption{\label{fpot}
The potential $U$ with false ($\phi_+$) and true vacuum
($\phi_-$).}
}
\end{figure}

As shown by Langer \cite{Langer67,Langer69}, the nucleation rate
$\Gamma$, that is the decay probability per time and per volume of a
metastable state represented by a local minimum of a potential, is
proportional to the imaginary part of a certain energy:
\be
\Gamma = - 2 \;\mathrm{Im} E \,.
\ee
The energy $E$ is given by the logarithm of a functional integral
\be
\mathcal{N} \int [d\phi] \, \e^{-\Hc(\phi)}\,,
\ee
with appropriate boundary conditions, where $\phi(x)$ is the local order
parameter and $\cal H$ is the Ginzburg-Landau Hamiltonian (a factor
$k_B T$ has been absorbed into $\cal H$).
The Hamiltonian is given by
\be
\Hc(\phi) = \int\!\!d^3x \left[ \frac{1}{2}
\left( \partial_{\mu} \phi(x) \right)^2 + U(\phi(x)) \right]\,,
\ee
with an asymmetric potential $U$.
We consider a potential of the type depicted in Fig.\ \ref{fpot}, with a
metastable minimum at $\phi=\phi_+$ and a stable minimum at
$\phi=\phi_-$. Following Coleman we call the phase corresponding to the
minimum at $\phi_+$ the false vacuum and the one corresponding to the
minimum at $\phi_-$ the true vacuum.

\begin{figure}[hbt]
\vspace{.8cm}
\centering
{\unitlength1cm
\begin{picture}(9,5)
\put(9,2.5){\vector(1,0){.1}}
\put(9.2,2.4){$r$}
\put(0,2.5){\vector(-1,0){.1}}
\put(-0.4,2.4){$r$}
\put(4.4,5.2){$\phi$}
\put(2.4,2){$R$}
\put(6.2,2){$R$}
\put(4.8,4.7){$\phi_+$}
\put(4.8,-.2){$\phi_-$}
\epsfig{file=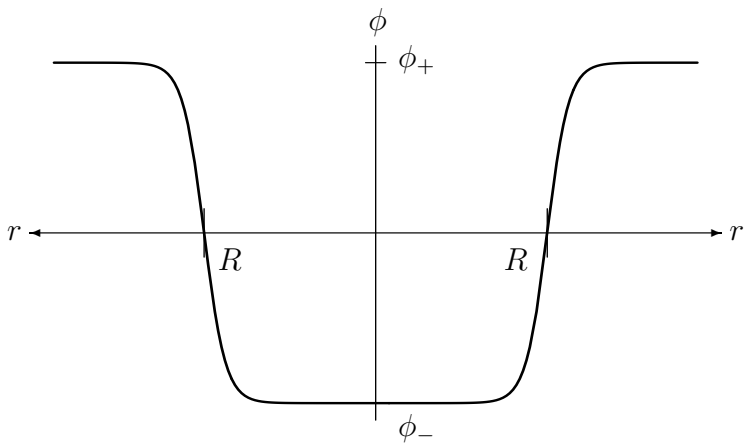,width=9cm,height=5cm}
\end{picture}
}
\parbox[t]{0.8\textwidth}{
\caption{\label{fbubble}
Profile of the critical bubble.}
}
\end{figure}

In a semiclassical approach the desired imaginary part of the functional
integral can be obtained by means of a saddle point expansion around the
classical solution which corresponds to the transition from the false to
the true vacuum \cite{Langer67,Coleman,Callan}.
This solution describes a critical bubble. A bubble centered at the
origin is represented by a radial symmetric function $\phi_c(r)$, which
depends on $r = \sqrt{x_{\mu} x_{\mu}}$ only. The boundary condition at
infinity
\be
\lim_{r \rightarrow \infty} \phi_c(r) = \phi_+
\ee
reflects that there is false vacuum outside the bubble.
The presence of the true vacuum inside the bubble means that the value
of the field at the center is near $\phi_-$:
\be
\phi_c(0) \approx \phi_- \,.
\ee
Due to differentiability of $\phi(x)$ we must have
\be
\left. \frac{d \phi_c}{dr}\right|_{r=0} = 0 \,.
\ee
The field equation for the bubble solution is
\be
\frac{d^2 \phi}{d r^2} + \frac{2}{r} \frac{d \phi}{dr} = U'(\phi) \,.
\ee
If we interpret $r$ as time and $\phi$ as the coordinate of a particle,
then this equation equals the equation of motion for a point particle in
the reversed potential $-U$ with a time-dependent friction term. From
the form of the potential it is intuitively clear that there is a unique
value of $\phi_c(0)$ near $\phi_-$, where the particle starts with zero
velocity, then rolls down the slope and climbs up the other hill to
approach its top $\phi_+$ asymptotically as time goes to infinity. In
fact, Coleman, Glaser and Martin \cite{CGM} have proved that such a
radial-symmetric non-trivial solution exists and that it is the one with
smallest energy apart from the trivial solution $\phi \equiv \phi_+$.
The qualitative form of the solution is as in Fig.\ \ref{fbubble}, which
shows a cross-section through the critical bubble.

For fields $\phi(x)$ near the classical solution $\phi_c$ the
Hamiltonian can be expanded up to quadratic terms as
\be
\Hc[\phi] = \Hc_c + \frac{1}{2} \int\!\!d^3 x
(\phi(x) - \phi_c(x)) \Mc (\phi(x) - \phi_c(x)) + \ldots \,,
\ee
with the energy of the critical bubble,
\be
\Hc_c = \Hc[\phi_c] \,,
\ee
and the fluctuation operator
\be
\Mc = - \partial^2 + U''(\phi_c(x)) \,.
\ee
Due to translation invariance the operator $\Mc$ has three zero modes
proportional to $\partial_{\mu} \phi_c(x)$. Furthermore, there is one
negative mode, which is related to the metastability of the false
vacuum. Namely, as has been indicated in the introduction, expansion or
contraction of the critical bubble lowers its energy, which means that
the corresponding mode belongs to a negative eigenvalue of $\Mc$. A
proof of the existence of a single negative mode under rather general
assumptions has been given by Coleman \cite{Coleman-mode}.

{}From the work of Langer and of Callan and Coleman it follows that the
functional integral under consideration acquires an imaginary part,
which is proportional to
\be
\left| \lambda_- {\det}'\Mc \right|^{-1/2} \e^{-\Hc_c}
\ee
in the Gaussian approximation, where $\det'$ is the determinant without
negative and zero modes, and $\lambda_-$ is the negative eigenvalue of
$\Mc$. If multibubble solutions are taken into account in a dilute gas
approximation, the final result for the nucleation rate is obtained as
\be
\label{Gamma}
\Gamma = \left( \frac{\Hc_c}{2 \pi} \right)^{3/2}
\frac{1}{\sqrt{|\lambda_-|}}
\left| \frac{\det'\Mc}{\det \Mcn} \right|^{-1/2}
\e^{-\Hc_c} \,.
\ee
Here the operator $\Mcn$ is the Helmholtz operator defined by
\be
\Mcn = - \partial^2 + U''(\phi_+) \,.
\ee
The expression (\ref{Gamma}) is of the form announced in the
introduction. In this work the above expression, in particular the
functional determinant, will be evaluated by field theoretic methods.
%
%
\section{The bubble solution}

We consider the standard Ginzburg-Landau potential consisting of a
symmetric double-well term,
\be
U_s = \frac{g}{4!}(\phi^2 - v^2)^2\,,
\ee
and an additional asymmetric term:
\be
\label{gpot}
U=U_s + \frac{\eta}{2 v}(\phi-v) + U_0 \,.
\ee
The constant
\be
U_0 = \frac{3 \eta^2}{8 v^4 g} + \frac{9 \eta^3}{16 v^8 g^2}
+ O(\eta^4)
\ee
is chosen such that $U(\phi_+)=0$ (cf.~Fig.~\ref{fpot}).
The parameter $\eta$ fixes the asymmetry of the potential. In
particular, the difference between the values of the potential at its
minima $\phi_{\pm}$ is
\be
U(\phi_+) - U(\phi_-) = 2 \eta + O(\eta^3)\,.
\ee
We define a bare mass parameter $m$ in terms of the symmetric part
$U_s$:
\be
m^2 =
\left. \frac{\partial^2}{\partial \phi^2} U_s(\phi) \right|_{\phi=v}
= \frac{gv^2}{3} \,.
\ee

The field equation for radially symmetric fields is
\be
\label{fieldeq}
- \frac{d^2 \phi}{d r^2} - \frac{2}{r} \frac{d \phi}{d r}
+ \frac{g}{6} \phi (\phi^2 - v^2) + \frac{\eta}{2v} = 0 \,.
\ee
Before we turn to a systematic approach to solving this equation let us
consider the thin wall approximation. Inspection of the field equation
in the light of the mechanical analogue mentioned in the introduction
shows that for small $\eta$ the solution is nearly equal to $-v$ inside
a sphere of radius $R$ and nearly equal to $+v$ outside. The region
where $\phi$ differs significantly from these values is a thin shell of
thickness $\theta$. The thin wall approximation amounts to
\be
\phi = \left\{
\begin{array}{rcl}
-v & \mbox{, for} & r < R - \theta/2 \\
\phi_k & \mbox{, for} & R - \theta/2 < r < R + \theta/2 \\
+v & \mbox{, for} & r > R - \theta/2 \,,
\end{array}
\right.
\ee
where the ``kink''
\be
\phi_k(r) = v \tanh \left( \frac{m}{2} (r-R) \right)
\ee
is a solution of
\be
- \frac{d^2 \phi}{d r^2} + \frac{g}{6} \phi (\phi^2 - v^2) = 0
\ee
and $\theta = 4 / m$.

The energy of a bubble solution can be written as
\be
\Hc_c = 4 \pi \int_0^{\infty}\!\!dr\,r^2
\left[ \frac{1}{2} \left( \frac{d \phi}{dr} \right)^2 + U_s (\phi) + U_0
\right]
+ \frac{2 \pi \eta}{v} \int_0^{\infty}\!\!dr\,r^2 (\phi - v) \,.
\ee
The second term is substantially different from zero only inside the
bubble. It yields the volume contribution
\be
\Hc_V = - \frac{4 \pi}{3} R^3 \eta \,.
\ee
The first term gets a substantial contribution only inside the wall,
\be
\Hc_S = 4 \pi \int_{R - \theta/2}^{R + \theta/2} dr\,r^2
\left[ \frac{1}{2} \left( \frac{d \phi}{dr} \right)^2 + U_s (\phi) + U_0
\right]
\approx 4 \pi R^2 \int\!\!dr \left( \frac{d \phi}{dr} \right)^2
= 4 \pi R^2 \sigma \,,
\ee
with
\be
\sigma = 2 \frac{m^3}{g} \,.
\ee
For small asymmetries $\eta$ the critical radius $R_c = 2 \sigma /
\eta$, for which the energy is stationary, gets large. Therefore the
wall is indeed thin compared to the size of the bubble. The total energy
in this approximation is
\be
\Hc_c = \frac{16 \pi \sigma^3}{3 \eta^2} \,.
\ee

For finite $\eta$ the solution of Eq.\ (\ref{fieldeq}) cannot be written
in closed form. In our approach the solution is constructed by means of
an expansion in powers of $\eta$. It is convenient to introduce
dimensionless variables
\be
\rt = \frac{m}{2} r , \hspace{5mm} \Rt = \frac{m}{2} R , \hspace{5mm}
\xi = \rt - \Rt , \hspace{5mm} \nt = \frac{g}{2 m^4} \eta ,
\ee
\be
\varphi (\xi) = \frac{1}{v} \phi (r) \,,
\ee
where the value of $\Rt$ will be fixed later. The field equation in
these variables is
\be
-\frac{d^2 \varphi}{d \xi^2}
- \frac{2}{\xi + \Rt} \, \frac{d \varphi}{d \xi}
+ 2 \varphi(\varphi^2 - 1) + \frac{4}{3} \nt =0\,.
\ee
Based on the thin wall approximation we write a Laurent series as an
ansatz for the critical radius,
\be
\label{RLaurent}
\Rt = \frac{a_{-1}}{\nt} + a_0 + a_1 \nt + a_2 \nt^2 + \dots \,,
\ee
and expand the field equation into powers of $\nt$:
\be
-\frac{d^2 \varphi}{d \xi^2}
- \frac{2}{a_{-1}} \, \nt \, \frac{d \varphi}{d \xi}
+ \nt^2 \, \frac{2 (\xi + a_0)}{a_{-1}^2} \, \frac{d \varphi}{d \xi}
+ 2 \varphi(\varphi^2 - 1) + \frac{4}{3} \nt + O(\nt^3) = 0 \,.
\ee
Its solution is obtained perturbatively up to second order by means of
the expansion
\be
\varphi = \varphi_0 + \nt \, \varphi_1 + \nt^2 \varphi_2 + O(\nt^3)\,.
\ee
To zeroeth order we get the well-known kink,
\be
\varphi_0(\xi) = \tanh(\xi)  \,.
\ee
The field equation to first order fixes the leading coefficient in
$\Rt$, Eq.\ (\ref{RLaurent}), as
\be
a_{-1}=1\,.
\ee
The first order solution obeying the correct boundary condition at $\xi
\rightarrow \infty$ reads
\be
\varphi_1 = -\frac{1}{3} - c \, \mbox{sech}^2\xi
\ee
with a free parameter $c$. The constant term reflects the shift of the
minimum
\be
\varphi_{\pm} = \pm 1 - \frac{\nt}{3} \mp \frac{\nt^2}{6}
- \frac{4\nt^3}{27} + O(\nt^4)\,.
\ee
The term proportional to $c$ can be traded against a shift in the
critical radius $\Rt$ in the lowest order solution according to
\be
\tanh(\xi - c \nt) = \tanh \xi - c \, \nt \, \mbox{sech}^2\xi
- c^2 \, \nt^2 \, \mbox{sech}^3\xi \, \sinh \xi + O(\nt^3)\,.
\ee
We can therefore set $c=0$ and remain with
\be
\varphi_1 = -\frac{1}{3}\,.
\ee
The equation to second order implies
\be
a_0 = 0
\ee
and has the solution
\bea
\varphi_2(\xi)&=&
- \frac{\xi}{2} (\tanh\xi - 1) + \frac{\xi}{6} (\cosh\xi-\sinh\xi)^2
- \frac{1}{2} \xi^2 \, \mbox{sech}^2\xi
- \frac{7}{12} \xi \, \mbox{sech}^2\xi
\nonumber\\
&&- \ln (1 + \e^{-2 \xi}) \left(\frac{1}{2} \xi \, \mbox{sech}^2\xi
+ \frac{1}{2} \tanh\xi \right)
- \frac{1}{3} \ln (1 + \e^{-2\xi}) \sinh\xi \, \cosh\xi \nonumber\\
&&- \frac{1}{12} \tanh\xi
+ \frac{1}{2} \, \mbox{sech}^2\xi \ T(\xi)\,,
\eea
where we define
\be
T(\xi) = \int_0^{\xi} \xi' \tanh\xi' d\xi' \,.
\ee
Whereas the first order solution only corresponds to shifts of the
minimum and of the critical radius, the second order solution describes
true deformations of the bubble. The boundary condition at $r=0$, i.e.
$\xi = -\Rt$, is fulfilled order by order in $\nt$. For example, the
leading order solution yields
\be
\varphi_{0}' (-\Rt) = \e^{- 2/\nt} (4 + O(\nt) ) \,,
\ee
which vanishes to all orders in $\nt$. Similar observations hold in
higher orders.

With the expression for $\varphi$ we can calculate the energy of a
bubble, which in dimensionless quantities is given by
\be
\Hc_c = \frac{12 \pi m}{g} \int_0^\infty\!\!d\rt \, \rt^2
\left\{ (\varphi'(\xi))^2 + \Big[ (\varphi^2(\xi) - 1)^2
- (\varphi_+^2 - 1)^2 \Big]
+ \left[ \frac{8}{3} \nt \, \Big( \varphi(\xi) - \varphi_+ \Big) \right]
\right\} \,.
\ee
The integrands are centered around the critical radius $\rt=\Rt$. The
integration range in $\xi$ can be extended to the whole real axis. The
error coming from this is proportional to factors of the type
$\e^{-\mbox{\scriptsize const.}/\nt}$ and vanishes to all orders in
$\nt.$

{}From the parity of the functions $\varphi_k (\xi)$ it follows that the
expression for the energy is of the form
\begin{displaymath}
\Hc_c = \frac{12 \pi m}{g}
(O_0 \Rt^2 + P_0) + \nt (L_1 \Rt + v_1 \Rt^3) + \nt^2 (O_2 \Rt^2 + P_2)
\end{displaymath}
\be
+\nt^3 (L_3 \Rt + v_3 \Rt^3) + O(\nt^4) \,.
\ee
An expression for the critical radius $\Rt$ is obtained from the
condition
\be
\frac{d\Hc_c}{d\Rt} \stackrel{!}{=} 0 \,.
\ee
Explicit calculation of the coefficients leads to two more terms in the
Laurent series for $\Rt$:
\be
\label{Rcrit}
\Rt= \frac{1}{\nt} + 0 + \frac{2 - 3 \pi^2}{36} \nt
+ 0 \cdot \nt^2 + O(\nt^3) \,,
\ee
so that the bubble is now completely determined to second order.
For the energy we get
\be
\label{Hcrit}
\Hc_c = \frac{12 \pi m}{g}
\left[ \frac{8}{9} \frac{1}{\nt^2} + \frac{2 (4 - 9 \pi^2)}{81}
+ O(\nt^2) \right]\,.
\ee
%
%
\section{The heat kernel of $\Mc$}

Our main task is to calculate the determinant ratio
\be
\frac{\det'\Mc}{\det \Mcn}\,,
\ee
which is part of the prefactor in the nucleation rate. In dimensionless
variables the corresponding operators are
\be
M = \frac{4}{m^2} \Mc
= - \frac{\partial^2}{\partial \xi_{\mu} \partial \xi_{\mu}}
+  6 \varphi^2 (\xi) - 2
\ee
and
\be
\Mn = \frac{4}{m^2} \Mcn
= - \frac{\partial^2}{\partial \xi_{\mu} \partial \xi_{\mu}}
+  6 \varphi_+^2 - 2 \,.
\ee
Substituting the bubble solution yields for the potential
\be
\label{Vxi}
V(\xi) = 6 \varphi^2 (\xi) - 2
= V_0(\xi) + \nt V_1(\xi) + \nt^2 V_2(\xi) + O(\nt^3)\,,
\ee
with the coefficients
\bea
V_0(\xi)&=&-6 \, \mbox{sech}^2 \xi + 4 \,,\\
V_1(\xi)&=&-4 \tanh \xi \,,\\
V_2(\xi)&=&\frac{2}{3} + \xi \left[ 4 \tanh\xi + 4 \sinh\xi \, \cosh\xi
- \left( 7 + 6 \ln 2 \right) \mbox{sech}^3\xi \, \sinh\xi \right]
\nonumber\\
&&- \tanh\xi \, \ln (\cosh\xi) \left( 6 \xi \, \mbox{sech}^2\xi
+ 4 \cosh\xi \, \sinh\xi + 6 \tanh\xi \right) \nonumber\\
&&- \left( 1 + 6 \ln 2 \right) \tanh^2\xi
- 4 \ln{2} \, \sinh^2\xi \nonumber\\
&&+ 6 \, \mbox{sech}^3\xi \, \sinh\xi \ T(\xi) \,.
\eea
For the free operator we find accordingly
\be
\Vn = 6 \varphi_+^2 - 2
= \Vn_0 + \nt \Vn_1 + \nt^2 \Vn_2 + O(\nt^3)\,,
\ee
with
\bea
\Vn_0&=&+ 4 \,,\\
\Vn_1&=&- 4 \,,\\
\Vn_2&=&- \frac{4}{3}\,.
\eea
The scaling dimensions of $\Mc$ and $\Mcn$ are equal, as we have checked
with the help of their zeta functions (see below). Therefore in the
determinant ratio the four removed eigenvalues lead to a supplementary
factor:
\be
\frac{\det'\Mc}{\det \Mcn} =
\left( \frac{4}{m^2} \right)^{4} \frac{\det'M}{\det \Mn} \,.
\ee
To get the same number of eigenvalues in the numerator and denominator,
we also remove the four lowest eigenvalues $\omn_0 = \Vn$ of the free
operator $\Mn$, which are equal to each other:
\be
\frac{\det'M}{\det \Mn} = (\omn_0)^{-4} \frac{\det'M}{\det' \Mn} \,.
\ee
The prime indicates the omission of the four lowest eigenvalues.
The remainder is written in Schwinger's proper time representation:
\be
\ln \frac{\det' M}{\det' \Mn}=
- \int_0^{\infty} \frac{dt}{t} \,
\mbox{Tr}' (\e^{-tM} - \e^{-t\Mn}) \,.
\ee
In the integrand we recognize the heat kernels $\exp (-tM)$ of the
operators $M$ and $\Mn$, respectively. The distribution of the large
eigenvalues determines the behaviour of the heat kernel for small $t$,
whereas the lowest eigenvalues determine the large-$t$ behaviour. The
strategy for the calculation of the integral over $t$ is to divide the
integration range into a small-$t$ part, where the heat kernel is
approximated by an asymptotic expansion, and a large-$t$ part, where the
low lying spectrum is employed \cite{DPY}.

Let us consider the small-$t$ region first. For small and positive $t$
an asymptotic expansion for the heat kernels, the socalled Seeley
expansion, exists \cite{Seeley}. For the trace of the heat kernels in
$D$ dimensions it is of the form
\be
\label{Seeley}
\mbox{Tr} (\e^{-tM} - \e^{-t\Mn})
= (4\pi t)^{-D/2} \sum_{n=1}^\infty  t^n \mathcal{O}_n \,.
\ee
There are various methods for the calculation of the coefficients
$\mathcal{O}_n$. Our calculation is based on the insertion of a plane
wave basis in the manner of \cite{DPY,Nepomechie}. By means of partial
integrations we managed to express the coefficients in terms of the
potential $V$ and the Laplacean $\partial^2$, which for a radial
symmetric potential like ours depends only on $\rt$ \cite{MR2}.
Inserting the potential $V(\xi)$, Eq.\ (\ref{Vxi}), and substituting the
expression for $\Rt$, Eq.\ (\ref{Rcrit}), we obtained with the help of
Mathematica \cite{Mathematica} the result
\begin{displaymath}
\mbox{Tr} (\e^{-tM} - \e^{-t\Mn}) =
\end{displaymath}
\begin{displaymath}
\frac{1}{(4 \pi t)^{3/2}} \left[ \frac{1}{\nt^2}
\left( \frac{112 \, \pi}{3} \, t
- \frac{160 \, \pi}{3} \, t^2 + \frac{832 \, \pi}{15} \, t^3
- \frac{11392 \, \pi}{315} \, t^4 + \frac{3328 \, \pi}{315} \, t^5
- \frac{49664 \, \pi}{24255} \, t^6 + \dots \right) \right.
\end{displaymath}
\begin{displaymath}
+ \left( -39.4801 \, t + 372.46 \, t^2 - 541.384 \, t^3 + 658.823 \, t^4
- 913.886 \, t^5 + 1225.92 \, t^6 - \dots \right)
\end{displaymath}
\be
\label{Seeleycoeff}
+ O(\nt^2)\bigg] \,.
\ee

{}From the leading term $t^{-1/2}$ it can be seen that the integral over
$t$ diverges at small $t$. It is well known that this divergence is
another disguise of the usual ultraviolet divergencies of quantum field
theory. They can be treated by means of dimensional regularization.
It is a peculiarity of the three-dimensional case that the regularized
expression is identical to the zeta-function regularized one without any
additional finite contribution \cite{Mue}. The zeta function of an
operator $A$ is in general defined by
\be
\zeta_A(z) = \mbox{Tr} A^{-z}
= \frac{1}{\Gamma(z)} \int_0^\infty\!\!dt \ t^{z-1} \,
\mbox{Tr} (\e^{- A t})
\ee
for sufficiently large Re\,$z$, where the integral converges, and
continued analytically to the rest of the complex plane. The
zeta-function regularized determinant is then given by
\be
\ln \det A = - \left. \frac{d}{dz} \zeta_A(z)\right|_{z=0} \,.
\ee
As we are interested in the ratio of two determinants, and want to
exclude zero and negative modes, we define
\be
\zeta^{\prime}(z) =
\frac{1}{\Gamma(z)} \int_0^\infty\!\!dt \ t^{z-1} \,
\mbox{Tr}' (\e^{- t M}-\e^{- t \Mn}) \qquad \mbox{for Re}\,z>1 \,.
\ee
It can be continued analytically to $z=0$ by separating the first term
in the Seeley expansion of $\mbox{Tr}(\e^{-tM}-\e^{-t\Mn})$,
Eqs.\ (\ref{Seeley}),(\ref{Seeleycoeff}):
\bea
\label{zetaprime}
\zeta^{\prime}(z)&=&
\frac{1}{\Gamma(z)} \int_0^\infty\!\!dt \ t^{z-1} \,
\left\{ \mbox{Tr}' (\e^{- t M} - \e^{- t \Mn})
- \Theta(1-t) \, \frac{1}{(4 \pi t)^{3/2}} \, t \, \mathcal{O}_1
\right\}\\
&&+ \frac{1}{(4 \pi )^{3/2}} \, \frac{1}{\Gamma(z)} \,
\frac{\mathcal{O}_1}{z - 1/2}\,.
\eea
Now the ratio of determinants can be expressed as
\be
\ln \frac{\det' M}{\det' \Mn}=
- \left. \frac{d}{dz} \zeta^{\prime}(z)\right|_{z=0} \,.
\ee

At this point let us check the relative dimensions of the two operators.
If the dimensions of $M$ and $\Mn$ differ by some number $u$ this
means that
\be
\frac{\det' \lambda M}{\det' \lambda \Mn} =
\lambda^u \frac{\det' M}{\det' \Mn} \,.
\ee
Representing the ratio of determinants by the derivative of the
zeta-function one obtains
\be
u = \zeta^{\prime}(0) \,.
\ee
In the expression (\ref{zetaprime}) for the analytically continued zeta
function the integral is convergent. Since $1/\Gamma(z)$ vanishes at
$z=0$, we find
\be
\zeta^{\prime}(0) = 0 \,,
\ee
so that the operators have equal dimensions as announced above.
%
%
\section{Calculation of the determinant}

As mentioned above, the strategy is to split the $t$-integration into a
small-$t$ part and a large-$t$ part. Therefore we introduce a parameter
$\Lambda$ separating the two regions, and split the zeta function as
\be
\zeta^{\prime}(z) = \zeta_<^{\prime}(z) + \zeta_>^{\prime}(z)
\ee
with
\be
\zeta_<^{\prime}(z) =
\frac{1}{\Gamma(z)} \int_0^\Lambda\!\!dt \ t^{z-1} \,
\mbox{Tr}' (\e^{- t M}-\e^{- t \Mn})
\ee
and $\zeta_>^{\prime}(z)$ correspondingly. In the same way the log of
the ratio of determinants is split as
\be
I \equiv \ln \frac{\det' M}{\det' \Mn} = I_< + I_>
\ee
with
\be
I_< = - \left. \frac{d}{dz} \zeta_<^{\prime}(z) \right|_{z=0} \,,
\ee
\be
I_> = - \left. \frac{d}{dz} \zeta_>^{\prime}(z) \right|_{z=0} \,.
\ee
Because the $t$-integration is not singular for $t > \Lambda$, we can
express $I_>$ directly as
\be
I_> = - \int_{\Lambda}^{\infty} \frac{dt}{t} \
\mbox{Tr}' (\e^{-tM} - \e^{-t\Mn}) \,.
\ee
%
%
\subsection{High-frequency part}

The behaviour of the heat kernel at small $t$ is governed by the high
frequencies in the spectrum of the operators. Therefore the contribution
to the determinant from the integration over small $t$ is its
high-frequency part. The small-$t$ expansion of the heat kernel is given
in Eq.\ (\ref{Seeleycoeff}). For the calculation of
$\zeta_<^{\prime}(z)$ we have to subtract the contributions of the
negative mode $\omega_- <0$ and the three zero-modes of $M$, and to add
the contribution of the four lowest eigenvalues of $\Mn$:
\be
\zeta_<^{\prime}(z) = \frac{1}{\Gamma(z)} \int_0^\Lambda\!\!dt \ t^{z-1}
\left\{ \frac{1}{(4 \pi t)^{3/2}} \sum_{n=1}^{\infty}
\mathcal{O}_n t^n - \e^{t |\omega_-|} - 3 + 4 \, \e^{-t\omn_0} \right\} \,.
\ee
For Re $z>1$ the integral can be performed:
\bea
\zeta_<^{\prime}(z)&=&\frac{1}{\Gamma(z)} \, \frac{1}{(4 \pi)^{3/2}}
\sum_{n=1}^{\infty}
\frac{\mathcal{O}_n} {z - 3/2 + n} \, \Lambda^{z - 3/2 + n}
\nonumber\\
&&- \frac{|\omega_-|^{-z}}{\Gamma(z)} \int_0^{|\omega_-| \Lambda} ds
\, s^{z-1} (\e^s - 1) - 4 \frac{\Lambda^z}{\Gamma(z+1)}
+ 4 \frac{(\omn_0)^{-z}}{\Gamma(z)} \, \gamma(z,\omn_0 \Lambda)\,,
\eea
where
\be
\gamma(a,x) = \int_0^x \e^{-t} \, t^{a-1} dt
\ee
is an incomplete gamma function \cite[sec.\ 6.5]{Abramowitz}.
This expression can be continued analytically to $z=0$, and the
derivative at this point yields
\bea
I_< &=& - \frac{1}{(4 \pi)^{3/2}} \sum_{n=1}^{\infty} \left\{
\frac{\mathcal{O}_n}{n - 3/2} \, \Lambda^{n - 3/2} \right\} \nonumber\\
&&+ \mbox{Ei}(|\omega_-| \Lambda) - \ln (|\omega_-| \Lambda) + 3 \gamma
+ 4 \mbox{E}_1 (\omn_0 \Lambda) + 4 \ln (\omn_0 \Lambda) \,,
\eea
where $\gamma = - \Gamma'(1) = 0.57721\ldots$ is Euler's constant and
\be
\mbox{Ei}(x) = - \mbox{P}\int_{-x}^{\infty} \frac{\e^{-t}}{t} dt \,,
\qquad x>0
\ee
\be
\mbox{E}_1(x) = \int_{x}^{\infty} \frac{\e^{-t}}{t} dt \,,
\qquad x>0
\ee
are the exponential integrals \cite[sec.\ 5.1]{Abramowitz}. Here the
explicit coefficients $\mathcal{O}_n$ from Eq.\ (\ref{Seeleycoeff}) and
the expressions for $\omega_-$ and $\omn_0$ given in the next section
are to be inserted.
%
%
\subsection{Low-frequency part}

The low eigenvalues of $M$ and $\Mn$ determine the behaviour of the heat
kernels for large $t$. We calculate the low-lying spectrum with the help
of a perturbative expansion in $\nt$.

In spherical coordinates the eigenvalue equation
\be
M v(\rt) = \omega v(\rt),
\ee
goes by means of the usual transformation
\be
v(\rt) = \frac{1}{\rt}\, \hat{\psi}(\rt)
\ee
and the shift
\be
\xi = \rt - \Rt, \qquad \psi(\xi) = \hat{\psi}(\rt)
\ee
over into
\be
\left[ - \frac{d^2}{d \xi^2} + \frac{l(l+1)}{(\xi + \Rt)^2} + V(\xi)
\right] \psi_{nl}(\xi) = \omega_{nl} \psi_{nl}(\xi)\,.
\ee
where $V(\xi)$ is given in Eq.\ (\ref{Vxi}), and $l = 0,1,2,\ldots$ is
the angular quantum number. With the help of Eq.\ (\ref{Rcrit}) the left
hand side is expanded in powers of $\nt$. To lowest order one finds the
P\"oschl--Teller potential $V_0=-6 \, \mbox{sech}^2\xi +4$, whose
eigenvalues are known exactly \cite{Rajaraman}. There exist two
discrete values
\bea
\omega_0^0&=&0\,, \qquad
\psi_0^0(\xi) = \sqrt{\frac{3}{4}} \, \mbox{sech}^2 \xi \,, \\
\omega_3^0&=&3\,, \qquad
\psi_3^0(\xi) = \sqrt{\frac{3}{2}}\,\sinh \xi \ \mbox{sech}^2 \xi \,,
\eea
and a continuum
\be
\omega_k^0 = k^2 + 4, \qquad k \in \mathbf{R} \,,
\ee
with the corresponding eigenfunctions
\be
\psi_k^0(\xi) \sim
\e^{i k\xi}(3 \tanh^2 \xi -1 -k^2 -3 i k \tanh \xi)\,.
\ee
Because of the radial symmetry, the problem is here only defined along
the half-axis $\rt>0$. The boundary condition at $\rt=0$, i.e.\
$\xi=-\Rt$, is obeyed up to terms which vanish to all orders in $\nt$.

To first order in $\nt$ the corrections to the eigenvalues vanish.

For the second order, we use the following trick \cite{Shankar}.
For every eigenfunction $\psi_n^0$ in the zeroeth order there is an
operator $\Omega_n$, fulfilling the relation
\be
[\Omega_n , (-\partial^2 + V_0)] \, \psi_n^0 = V_1 \psi_n^0 \,.
\ee
The second correction in the eigenvalue is then given by
\be
\omega_n^2 = \langle \psi_n^0 | V_1 \Omega_n + V_2 | \psi_n^0 \rangle
+ l(l+1) \,.
\ee
This way, we find the second order for the discrete eigenvalues:
\bea
\omega_0^2&=&l(l+1) - 2 \,,\\
\omega_3^2&=&l(l+1) + 3 - \pi^2 \,,
\eea
using
\be
\Omega_0 = \xi\,, \qquad
\Omega_3 = 2 \xi -\frac{\cosh \xi}{\sinh \xi} \,.
\ee
So the discrete eigenvalues of $M$ are
\bea
\omega_{0l}&=&\nt^2 \big( l(l+1) - 2 \big) + O(\nt^4) \\
\omega_{3l}&=& 3 + \nt^2 \big( l(l+1) + 3 - \pi^2 \big) + O(\nt^4) \,.
\eea
They are $(2l+1)$-fold degenerate.

In particular, the negative mode is given by
\be
\label{negmode}
\omega_- = \omega_{00} = - 2 \nt^2 + O(\nt^4) \,,
\ee
and the three zero modes are $\omega_{01}$.

The band of eigenvalues $\omega_{0l}$ near zero gives a contribution to
$I_>$ which reads
\be
I_>^0 (\Lambda) = - \int_{\Lambda}^{\infty} \frac{dt}{t}
\sum_{l=2}^\infty (2l+1) \e^{- \omega_{0l} t} \,.
\ee
This expression can be evaluated with the help of various nontrivial
relations involving special functions. The result is
\be
I_>^0(\Lambda) = - \frac{1}{\nt^2 \Lambda}
- \frac{5}{3} \ln (\nt^2 \Lambda) + c_0 + O(\nt^2 \Lambda)
\ee
with
\be
c_0 = \frac{9}{2} - \ln 54 - 4 \zeta_{R}'(-1) - \frac{5}{3} \gamma
= 0.21068\ldots \,.
\ee
However, the corrections to the eigenvalues from next order perturbation
theory would produce corrections of order $(\nt^2\Lambda)^0$, and
therefore we write
\be
I_>^0(\Lambda) = - \frac{1}{\nt^2 \Lambda}
- \frac{5}{3} \ln (\nt^2 \Lambda) + O((\nt^2 \Lambda)^0) \,.
\ee
In a similar way the band of eigenvalues $\omega_{3l}$ near three gives
a contribution
\be
I_>^3(\Lambda) = - \frac{3}{\nt^2} \Gamma(-1, 3 \Lambda)
+ O(\nt^0) \,.
\ee
with the incomplete gamma function \cite[sec.\ 6.5]{Abramowitz})
\be
\Gamma(a,x) = \int_x^{\infty}\!\e^{-t} \, t^{a-1} dt \,.
\ee
The term of order $\nt^0$ has been calculated, but is not displayed,
because $I_>^0$ already contains an uncertainty of this order.

Now we turn to the remaining spectrum. The continuous eigenvalues of $M$
and all eigenvalues of $\Mn$ can be written in the form
\be
\omega_{kl} = k^2 + \Vn + \nt^2 l(l+1) + O(\nt^4) \,, \qquad k \geq 0\,,
\ee
with
\be
\Vn = 4 - 4\nt -\frac{4}{3} \nt^2 + O(\nt^3) \,.
\ee
In order to calculate their contribution to the heat kernel one needs
the difference of the spectral densities $\varrho_l(k)$ and
$\varrho^{(0)}_l(k)$. We have calculated the spectral densities in the
framework of perturbation theory. This was done by extracting the phase
shifts $\delta_l(k)$ from the asymptotic behaviour of the wavefunctions
and using the relation
\be
\varrho_l(k) - \varrho^{(0)}_l(k)
= \frac{1}{\pi} \frac{\partial \delta_l}{\partial k} \,.
\ee
It turns out that in $n$-th order of perturbation theory there are
terms proportional to $(\nt \Rt)^n$, which contribute to the lowest
order in $\nt$, because $\Rt \sim 1/\nt$. Summing up all these terms we
have been able to obtain the spectral densities to lowest order only.
Omitting the details, the contribution of the continuous spectra to
the trace of the heat kernels (leaving out the four lowest eigenvalues
of $\Mn$) is given by
\be
\sum_{l=0}^\infty (2l+1)
\int_0^\infty\!\!dk \, \Big( \varrho_l(k) - \varrho^{(0)}_l(k) \Big)
\e^{- \omega_{kl} t}\ + 4 \e^{-t \omn} =
\ee
\be
- \frac{1}{\nt^2 \, t} \left[ \e^{-3 t} + 1 - \left( \e^{-3 t}
\Phi(\sqrt{t}) + \Phi(\sqrt{4 t}) \right)
- \frac{2}{3} \Vn_1 \sqrt{\frac{t}{4\pi}} \, \e^{-4 t} \right]
+ O(\nt^0) \,,
\ee
where
\be
\Phi(x) = \frac{2}{\sqrt{\pi}} \int_0^{x}\!\!\e^{-t^2} dt
\ee
is the error integral.
{}From this we get the contribution to $I_>$ by integration over $t$:
\begin{displaymath}
I_>^k(\Lambda) =
\frac{1}{\nt^2} \left[ 3 \Gamma(-1,3 \Lambda)
- \frac{4}{3 \sqrt{\pi}} \Gamma(-1/2, 4 \Lambda)
+ \frac{1}{\Lambda \sqrt{\pi}} \Gamma(1/2, \Lambda)
- \int_{\Lambda}^\infty \frac{dt}{t^2}\ \Phi(\sqrt{t}) \e^{-3t} \right]
\end{displaymath}
\be
+ O(\nt^0) \,.
\ee
We now have all contributions to the low-frequency part $I_> = I_>^0 +
I_>^3 + I_>^k$ available. In Fig.\ \ref{ficut} they are presented as a
function of $\Lambda$ for $\nt=0.1$. One can see that for
$\Lambda > 1$ the contribution $I_>^0$ from the band near zero dominates
the sum.
%
%
\subsection{Composition of the determinant}

With all pieces at hand the logarithm of the determinant
\be
I = \ln \frac{\det' M}{\det' \Mn} = I_<(\Lambda) + I_>(\Lambda)
\ee
can now be composed.
The uncertainty in our result for $I_<$ is of order $\nt^2$. On the
other hand, for $I_>$ there are already unknown contributions of order
$\nt^0$. In the numerical evaluations we nevertheless include the
constant terms from our calculation of $I_>^0$ and $I_>^3$. Since the
result for $I$ is strongly dominated by $I_<$, as will be seen below,
the influence of the unknown corrections is expected to be
numerically small.

\begin{figure}[hbt]
\vspace{.8cm}
\centering
{\unitlength1cm
\begin{picture}(9,5)
\put(9.1,0.05){$t$}
\put(0.6,5.1){$K'$}
\epsfig{file=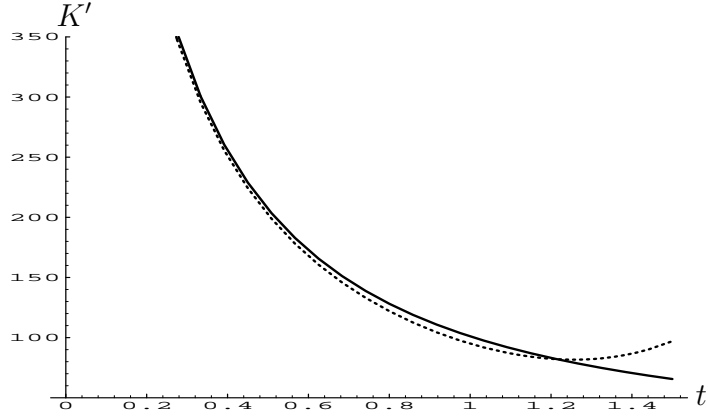,width=9cm,height=5cm}
\end{picture}
}
\parbox[t]{0.8\textwidth}{
\caption{\label{fgros}
The heat kernel $K'$ as a function of $t$ for $\nt=0.1$. The
dotted curve represents the Seeley expansion and the full curve is the
approximation from summing over the eigenvalues.}
}
\end{figure}

In an exact calculation $I$ would be independent of the artificial
cutoff parameter $\Lambda$. Using the approximative expressions above, a
dependence on $\Lambda$ of course shows up, and we have to make a good
choice. As a first guide we consider the heat kernel calculated from the
Seeley expansion, Eq.\ (\ref{Seeley}), and from the eigenvalues,
respectively. In Fig.\ \ref{fgros} the kernel
\be
K'(t) = \mbox{Tr}' (\e^{- t M}-\e^{- t \Mn})
\ee
is shown as function of $t$ for $\nt=0.1$ from the two approximations.

The small-$t$ approximation and the large-$t$ approximation are in good
agreement for $t<1.3$. A value of $\Lambda$ near 1.2, where the two
curves intersect each other, appears to be reasonable for this value of
$\nt$.

\begin{figure}[hbt]
\vspace{.8cm}
\centering
{\unitlength1cm
\begin{picture}(9,5)
\put(1.1,5.2){$I$}
\put(9.2,1.6){$\Lambda$}
\epsfig{file=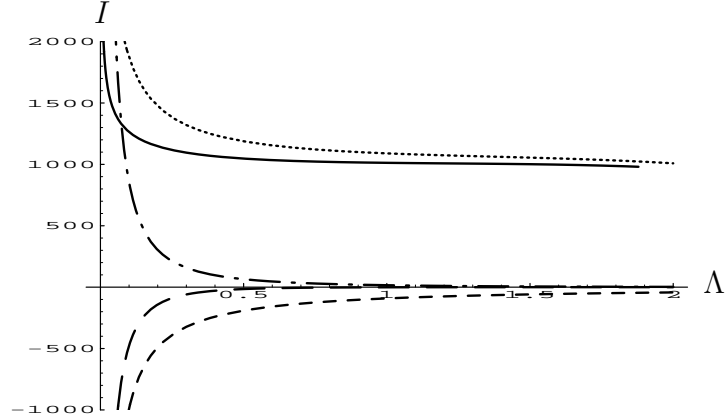,width=9cm,height=5cm}
\end{picture}
}
\parbox[t]{0.8\textwidth}{
\caption{\label{ficut}
The different contributions to $I$ as a function of $\Lambda$
for $\nt=0.1$:
$I_<$ ($\cdots\cdots$), \mbox{$I_>^0$(-- -- -- )}, $I_>^3$( ---~---),
$I_>^k$(-- $\cdot$ -- $\cdot$).
The full curve is the sum of the four contributions.}
}
\end{figure}

To make things more quantitative, the various contributions to $I$ and
the total sum are shown as a function of $\Lambda$ in Fig.\ \ref{ficut}.
A broad plateau, where the $\Lambda$-dependence is rather small can
clearly be recognized.

An optimal value for $\Lambda$ can be determined in the following way.
As discussed above, the expression for $I$ can be expanded in powers of
$\nt$. The coefficients are functions of $\Lambda$. Requiring
\be
\frac{dI}{d\Lambda} = 0
\ee
order by order in $\nt$ leads to
\be
\Lambda = 1.11672 + 17.3175 \, \nt^2 + O(\nt^3) \,.
\ee
With this value one obtains
\be
I(\nt) = \frac{10.158}{\nt^2} - \frac{5}{3} \ln \nt^2 - 13.63 +
O(\nt^0)
\ee
with an error estimate of
\be
\Delta I(\nt) = \frac{0.0527}{\nt^2} - 6.37 + O(\nt^0) \,.
\ee
The systematic uncertainties of order $\nt^0$ come from the
corresponding unknown terms in $I_>$. Their contribution should be
numerically small, as can be seen from Fig.\ \ref{ficut}. The error
$\Delta I$ is estimated from the error of the Seeley expansion, which in
turn is taken to be given by its highest term.
%
%
\subsection{Improved calculation of the determinant}

The result for the determinant can be improved in various ways. First of
all, one observes that the high frequency (small $t$) part dominates the
result. Also, the low frequency part contains uncertainties of constant
order in $\nt$. Therefore it is desirable to use as much information as
possible from the high frequency part, i.e., from the Seeley expansion.

The determinant could be estimated from the Seeley expansion alone by
introducing a smooth exponential cutoff \cite{Carson}. With a free
cutoff parameter $\mu$ one writes
\be
K'(t) = \mbox{Tr}' (\e^{- t M}-\e^{- t \Mn})
= \e^{-\mu t} \mbox{Tr}' (\e^{- t M}-\e^{- t \Mn}) \, \e^{\mu t}\,,
\ee
and expands the modified kernel as
\be
\mbox{Tr}' (\e^{- t M} - \e^{- t \Mn}) \, \e^{\mu t}
= \frac{1}{(4\pi t)^{3/2}} \sum_n h_n(\mu) t^n + \sum_n g_n(\mu) t^n \,,
\ee
where the coefficients $h_n$ come from the Seeley expansion
(\ref{Seeleycoeff}) and the coefficients $g_n$ from the subtracted
negative mode and zero modes. Due to the factor $\exp (-\mu t)$ the
$t$-integration can be extended to infinity. In this way one gets an
estimate for $I$ from the small-$t$ expansion alone.

We have used this method and obtained results which are in fair
agreement with the earlier ones. It is, however, possible to introduce
a further improvement. From the knowledge of the band of low-lying
eigenvalues $\omega_{0l}$ it is possible to calculate its contribution
to $I$ completely \cite{exactI}. The outcome is
\bea
I^0 &=&- \frac{5}{3} \ln \nt^2
+ \frac{9}{2} - \ln 54 - 4 \zeta_{R}'(-1) + O(\nt^2) \\
&=&- \frac{5}{3} \ln \nt^2 + 1.172\,700\,5 + O(\nt^2)\,.
\eea
This result can be employed in the calculation of $I$ by separating the
contribution of this band of eigenvalues. This means that from the
Seeley expansion above the small-$t$ expansion of
\be
\sum_{l=2}^\infty (2l+1) \e^{- \omega_{0l} t}
= \frac{1}{\nt^2 t} - \frac{5}{3} + O(\nt^2 t)
\ee
is subtracted and the rest is treated according to the exponential
cutoff method. To the result the expression for $I^0$ is added finally.

The value of the cutoff-parameter $\mu$ has been obtained with the same
procedure as in the case of $\Lambda$ by requiring $dI/d\mu$ to vanish.
For the log of the determinant we get in this way
\be
I(\nt) = \frac{10.037}{\nt^2} - \frac{5}{3} \ln \nt^2 - 7.98
- 4 \, \nt + O(\nt^2) \,.
\ee
We have estimated the error by means of the Shanks extrapolation and
error determination \cite{Shanks} and obtained
\be
\Delta I(\nt) = \frac{0.00154}{\nt^2} - 0.084 - 0.795 \, \nt
+ O(\nt^2) \,.
\ee

A last improvement is based on the knowledge of the exact leading terms
in the small-$\nt$ expansion of $I(\nt)$ \cite{exactI}. From the results
for the discrete and continuous spectra of $\Mn$ it is possible to
derive
\be
I = \frac{c}{\nt^2} - \frac{5}{3} \ln \nt^2 + O(\nt^0) \,,
\ee
with
\be
c = \frac{20}{3} + 3 \ln 3 = 9.9625 \,.
\ee
So we write our final result for the determinants as
\be
\ln \frac{\det' M}{\det \Mn} = I - 4 \ln \omn_0
= \frac{c}{\nt^2} - \frac{5}{3} \ln \nt^2 - 13.52 + O(\nt^2) \,.
\ee
%
%
\section{The nucleation rate}

Having obtained the determinant of the fluctuation operator the
nucleation rate can be calculated according to Eq.\ (\ref{Gamma}), where
the energy of the critical bubble, Eq.\ (\ref{Hcrit}), and the negative
mode, Eq.\ (\ref{negmode}), have to be inserted. In terms of the mass
$m$ and the dimensionless parameters $\nt$ and
\be
u = \frac{g}{m}
\ee
the nucleation rate is
\begin{displaymath}
\Gamma =
\frac{m^3}{u^{3/2} \, \nt^{7/3}}
\exp\left[-\left( \frac{32\pi}{3}\,\frac{1}{u} + \frac{c}{2} + O(u)
\right) \frac{1}{\nt^2}
\right.
\end{displaymath}
\be
\left.
+ \left( \frac{8\pi}{27} (9\pi^2 - 4) \frac{1}{u} + 6.845 + O(u) \right)
+ O(\nt^2) \right]\,.
\ee
The parameters appearing in the Hamiltonian are, however, not
immediately accessible in phenomenological applications or in a field
theoretical context. More appropriate are the renormalised parameters,
which are directly related to measurable quantities.
Even more important in this context is the fact that in the dimensional
regularisation scheme around $d=3$ dimensions the divergencies in the
relation between bare and renormalised parameters are not visible in the
one-loop approximation. Therefore physical quantities should be expressed in
terms of renormalised parameters.

We shall use the renormalised quantities as, e.g., specified in
\cite{Mue}. The renormalised mass $m_R$ and the field renormalisation
constant $Z$ are defined in terms of the inverse propagator at small
momenta:
\be
G^{-1}(p) = \frac{1}{Z} \Big( m_R^2 + p^2 + O(p^4) \Big) \,.
\ee
The renormalised mass is equal to the inverse of the second moment
correlation length
\be
\xi^{(2)} = \frac{1}{m_R} \,.
\ee
The renormalised field is given by
\be
\phi_R = Z^{-1/2} \phi
\ee
and the renormalised field expectation value
\be
v_R = Z^{-1/2} (v + \langle \phi \rangle)
\ee
correspondingly. The renormalised coupling, defined in terms of the
mass and the field expectation value by
\be
g_R = \frac{3 m_R^2}{v_R^2}\,,
\ee
has dimensions of a mass. Its dimensionless counterpart is
\be
u_R=\frac{g_R}{m_{R}}\,.
\ee
The renormalised dimensionless asymmetry parameter is
\be
\nt_R = \frac{g_R}{2 m_R^4} \eta\,.
\ee
Up to first order, the relations between the bare and the renormalised
quantities are given by \cite{MH}:
\bea
m&=&m_R \left\{ 1 - \frac{3}{128 \pi} u_R + O(u_R^2) \right\} \,,
\nonumber\\
u&=&u_R \left\{ 1 + \frac{31}{128 \pi} u_R + O(u_R^2) \right\}\,.
\eea
Expressed in terms of the renormalised parameters the nucleation rate is
\begin{displaymath}
\Gamma =
\frac{m_R^3}{u_R^{3/2} \, \nt_R^{7/3}}
\exp\left[-\left( \frac{32\pi}{3}\,\frac{1}{u_R} + \frac{c}{2} -
\frac{113}{12} + O(u_R) \right) \frac{1}{\nt_R^2}
\right.
\end{displaymath}
\be
\left.
+ \left( \frac{8\pi}{27} (9\pi^2 - 4) \frac{1}{u_R} + 0.758 + O(u_R)
\right) + O(\nt_R^2) \right]\,.
\ee

This formula is an analytical expression for $\Gamma$ which for the first
time includes a complete treatment of quadratic fluctuations.
Compared to the thin wall approximation,
\be
\Gamma_{\mathrm{TWA}} =
\exp\left[- \frac{32\pi}{3}\,\frac{1}{u_R \, \nt_R^2} \right]\,,
\ee
the leading term for small asymmetries $\eta$,
\be
\Gamma =
\frac{m_R^3}{u_R^{3/2} \nt_R^{7/3}}
\exp\left[-\left( \frac{32\pi}{3}\,\frac{1}{u_R} + \frac{c}{2} -
\frac{113}{12} \right) \frac{1}{\nt_R^2} \right]\,,
\ee
completes the thin wall approximation by giving the prefactor in
addition to the energy of the critical bubble.

The next-to-leading terms go beyond the thin wall approximation.
They can be employed to obtain an estimate for the region of validity of
the small $\nt_R$ expansion. Comparing the terms proportional to
$u_R^{-1}$ in the exponential and requiring the correction to be smaller
than the leading term we get
\be
\nt_R < 0.65\,.
\ee

The result for $\Gamma$ can be used to obtain estimates for the
nucleation rate by inserting phenomenological or measured values for the
physical parameters.
Moreover it can be employed in the context of field theory for an estimate
of the decay of a false vacuum at high temperatures.

Baacke and Kiselev \cite{Baacke} have calculated the nucleation rate in
three-dimensional scalar field theory by evaluating the fluctuation
determinant numerically. Our results can, however, not be compared directly,
because the authors of \cite{Baacke} pick out the terms relevant in a
certain high temperature limit. Also the renormalization scheme is
different from ours.

Another numerical calculation of the nucleation rate in the framework of
renormalisation group improved effective average actions has been
presented in \cite{Strumia}. In this case, too, it is not possible
without further information to compare their results with ours, because
the results of \cite{Strumia} are expressed in terms of parameters,
whose relation to the renormalised parameters used here is not clear.


%
\end{document}